\documentstyle[11pt,emulateapj5,natbib]{aastex}
\newcommand{\CII}{\hbox{{\rm C}\kern 0.1em{\sc ii}}}
\newcommand{\CIV}{\hbox{{\rm C}\kern 0.1em{\sc iv}}}

\newcommand{\FeI}{\hbox{{\rm Fe}\kern 0.1em{\sc i}}}
\newcommand{\FeII}{\hbox{{\rm Fe}\kern 0.1em{\sc ii}}}
\newcommand{\SiII}{\hbox{{\rm Si}\kern 0.1em{\sc ii}}}
\newcommand{\AlII}{\hbox{{\rm Al}\kern 0.1em{\sc ii}}}
\newcommand{\NiII}{\hbox{{\rm Ni}\kern 0.1em{\sc ii}}}
\newcommand{\CrII}{\hbox{{\rm Cr}\kern 0.1em{\sc ii}}}

\newcommand{\ZnII}{\hbox{{\rm Zn}\kern 0.1em{\sc ii}}}
\newcommand{\NII}{\hbox{{\rm N}\kern 0.1em{\sc ii}}}
\newcommand{\OIII}{\hbox{{\rm O}\kern 0.1em{\sc iii}}}
\newcommand{\OII}{\hbox{{\rm O}\kern 0.1em{\sc ii}}}
\newcommand{\OI}{\hbox{{\rm O}\kern 0.1em{\sc i}}}
\newcommand{\MgI}{\hbox{{\rm Mg}\kern 0.1em{\sc i}}}
\newcommand{\MgII}{\hbox{{\rm Mg}\kern 0.1em{\sc ii}}}

\newcommand{\HI}{\hbox{{\rm H}\kern 0.1em{\sc i}}}
\newcommand{\HII}{\hbox{{\rm H}\kern 0.1em{\sc ii}}}
\newcommand{\lya}{\hbox{{\rm Ly}\kern 0.1em$\alpha$}}
\newcommand{\Ly}{\hbox{{\rm Ly}\kern 0.1em$\alpha$}}
\newcommand{\Ha}{\hbox{{\rm H}\kern 0.1em$\alpha$}}
\newcommand{\Hb}{\hbox{{\rm H}\kern 0.1em$\beta$}}

\newcommand{\flux}{erg~s$^{-1}$~cm$^{-2}$}

\newcommand{\mpy}{\hbox{$M_{\odot}$~yr$^{-1}$}}

\newcommand{\msun}{\hbox{$M_{\odot}$}}
\newcommand{\cmsq}{\hbox{cm$^{-2}$}}
\newcommand{\NHI}{\hbox{$N_{\HI}$}}

\newcommand{\kpc}{\hbox{$h^{-1}$~kpc}}

\newcommand{\kms}{km~s$^{-1}$}

\newcommand{\EW}{\hbox{$W_{\rm r}^{\lambda2796}$}}

\newcommand{\nfield}{21}		%%z=1
\newcommand{\ndetected}{14}	        %%z=1

\begin{document}

%Discovering and measuring the kinematics of $z\sim1$--$2$ analogs to M82 with SINFONI~\altaffilmark{1}\\OR\\
%The kinematics of $z\sim1$  analogs to M82 with SINFONI~\altaffilmark{1}\\OR\\
%The SINFONI \MgII\ Program: Early Results\\OR\\
\title{The SINFONI \MgII\ Program for Line Emitters (SIMPLE): discovering starbursts near QSO sightlines\altaffilmark{1}
}%end title

\author{Nicolas Bouch\'e\altaffilmark{2},  Michael T. Murphy\altaffilmark{3}, C\'eline P\'eroux\altaffilmark{4}, 
Richard Davies\altaffilmark{2}, Frank Eisenhauer\altaffilmark{2}, Natascha M. F\"orster Schreiber\altaffilmark{2}, Linda
Tacconi\altaffilmark{2}}

\altaffiltext{1}{Based on observations made at the ESO telescopes under
program ID 077.A-0576, 078.A-0600, 078.A-0718 and 079.A-0600.}
\altaffiltext{2}{Max Planck Institut f\" ur  extraterrestrische Physik, Giessenbachstrasse, D-85748 Garching, Germany;
NB: nbouche@mpe.mpg.de}
\altaffiltext{3}{Centre for Astrophysics \&\ Supercomputing, Swinburne University of Technology, 
 Hawthorn, Victoria 3122, Australia}
\altaffiltext{4}{European Southern Observatory, Karl-Schwarzschild-str 2, D-85748 Garching, Germany}

\slugcomment{Received 2007 April 28; accepted 2007 September 12; to appear in ApJL 2007 November 1} 
 
 
%%%%%%%%%%%%%%%%%%%%   ABSTRACT  %%%\%%%%%%%%%%%%%%%%%%%%%
\begin{abstract}
Low-ionization transitions such as the \MgII\ $\lambda$2796/2803 doublet 
trace cold gas in the vicinity of galaxies. It is not clear whether
this gas is part of the interstellar medium
of large proto-disks, part of dwarfs, or part of entrained material in supernova-driven outflows.
Studies based on \MgII\ statistics, e.g.  stacked images  and clustering analysis,
 have  invoked starburst-driven outflows where \MgII\ absorbers are tracing the denser and colder gas of the outflow.
 A consequence of the outflow scenario is that the strongest absorbers
ought to be associated with starbursts. 
We use the near-IR integral field spectrograph SINFONI to test whether starbursts are found around  
  $z\sim1$ \MgII\ absorbers.
For 67\%\  (\ndetected\ out of \nfield) of the absorbers with rest-frame equivalent width $\EW>2$~\AA, 
 we do detect \Ha\ in emission within $\pm200$~\kms\ of the predicted wavelength based on the \MgII\ redshift,
 and with impact parameter ranging from 0.2\arcsec\ to 6.7\arcsec\ from the QSO.
 The star formation rate (SFR) inferred from \Ha\ is $1$--$20$~\mpy, i.e.
 showing a level of star formation larger than in M82 by a factor of $>4$ on average.
Our flux limit (3-$\sigma$) is $f_{\Ha}<1.2\times10^{-17}$~\flux, corresponding to a SFR of $\sim$0.5~\mpy,
 much below past ground-based \Ha\ surveys of absorption-selected galaxies.
We find evidence (at $>95$\%) for a correlation between SFR and equivalent width, 
indicating a physical connection between
starburst phenomena and gas seen in absorption.
In the cases where we can extract the velocity field, the  host galaxies of \MgII\ absorbers with $\EW>2$~\AA\
reside in halos with  mean mass  $<\log M_h(\msun)>\sim 11.2$ in good agreement
with clustering measurements.
\end{abstract}


%\tableofcontents

%%%%%%%%%%%%%%%%%%%%%%%%% INTRODUCTION %%%%%%%%%%%%%%%%%%%%%%%
\section{Introduction}

In the era of large spectroscopic surveys such as SDSS~\citep{YorkD_00a},
the \MgII\ $\lambda$2796/2803 doublet seen in background QSO spectra 
is particularly useful for selecting thousands of absorbing galaxies \citep[e.g.][]{YorkD_06a}.
Galaxies selected via their \MgII\ $\lambda$2796/2803 doublet  
have been used as proxies for studying \HI\ selected galaxies at $z<1.5$
\citep[e.g.][]{RaoS_06a}.

It is not clear whether this cold gas is part of the interstellar medium of large spirals \citep{WolfeA_86a},
part of the halos of galaxies \citep{BahcallJ_69a} 
or part of dwarf galaxies \citep{YorkD_86a} akin to  the Magellanic Clouds. An alternative
scenario for absorption-selected galaxies is that the gas
seen in absorption is  part of cold gas clumps in the host galaxy halos
 entrained in outflows produced by supernovae (SNe)  \citep{NulsenP_98a,SchayeJ_01a}.
  Most likely all three scenarios play a role,
but in what proportion for a given equivalent width or \HI\ column density?

Using cross-correlation techniques,
 \citet{BoucheN_06c} statistically constrained the halo mass of $\sim$2000 \MgII\ host galaxies,   finding that
the host halo mass ($M_h$) decreases with increasing \MgII\ rest-equivalent width (\EW). Given that the equivalent width
must be correlated with the line-of-sight velocity width, $\Delta v$, 
as also verified observationally by \citet{EllisonS_06a},
these results imply that $M_h$ and $\Delta v$ are anti-correlated. This seems to show that the clouds responsible
for the absorption are not virialized, otherwise, a $M_h$--$\Delta v$ correlation would 
 have been measured.  A natural explanation   for this is  that  the cold gas originates
  from SN-driven outflows, in which case the velocity width $\Delta v$
may be related to the outflow kinematics or may be a measure of the mass outflow rate.
In a few cases,   evidence for super-winds  is seen  from the absorption profile \citep{BondN_01a,EllisonS_03a}.
Others, e.g. \citet{ProchterG_05a} and \citet{MartinC_06a} arrived at the same conclusion 
that super-winds play a significant role in \MgII\ absorbers from  very different perspectives.

 The most important implication from the starburst scenario is that 
  the selection of galaxies via the \MgII\ absorption signature 
 may be equivalent to selecting starburst-producing super-winds.
 Equivalently, non-detections of the \Ha\ signature of a starburst
 would enable one to rule out the starburst scenario,
 while the presence of \Ha\ emission is no un-ambiguous proof of starburst-driven winds.

In this letter, we report the first results  of our ``SINFONI \MgII\ Program for Line Emitters'' (SIMPLE),
which is aimed at detecting \Ha\ from the starburst galaxy   using
 the integral field unit (IFU)   SINFONI  \citep{EisenhauerF_03a} available at the Very Large Telescope (VLT).
 The advantages of IFUs include the possibility of detecting the \MgII\ host at impact parameters smaller than
the seeing disk, and of measuring the  two-dimensional kinematics
of galaxies.

We present our sample selection in section~2, and results in section~3. Throughout, we
use the $h=0.7$, $\Omega_M=0.3$, $\Omega_\Lambda=0.7$ cosmology.

\section{Sample Selection}

 Our sample of \MgII-selected galaxies 
is selected  from the SDSS/2QZ databases and
from the compilation of \citet{RyabinkovA_03a}
 with the only physical criterion $\EW>2$~\AA. We then ensured
 that the corresponding \Ha\ emission line would fall inside the SINFONI wavelength range 
  and away from sky OH emission lines.
The $\EW>2$~\AA\ criterion was used for the following two reasons.
First, in the super-wind scenario, the strongest absorbers (as measured by \EW) ought
to have the largest star-formation rates, hence the largest \Ha\ fluxes.
Second, the $\EW>2$~\AA\ criterion is known to select the hosts with the smallest impact parameters,
$\rho<30$~\kpc\ \citep[e.g.][]{SteidelC_95b,BoucheN_06c}.
At $z\simeq1$, this corresponds to $\sim4$\arcsec, which means that the host galaxy will fall within
the field of view of SINFONI (8\arcsec).

We have obtained SINFONI observations towards \nfield\ \MgII\ absorbing galaxies all with
absorption redshift $z_{\rm abs}\simeq1$, i.e. corresponding the $J$-band filter.
We focused on $z\sim1$ systems because (1) this redshift range is close to the range used in
 \citet{BoucheN_06c},
and (2) our flux sensitivity is enhanced in the J-band  compared to that in the K-band.
The latter point is due to the gain in luminosity distance compared to the loss in raw instrument throughput.
For each QSO field  we took  4$\times$600s exposures, placing the QSO in the four quadrants of the field of view.
The total exposure time
 for the host galaxy thus varies between 10 and 40 minutes, depending
on its location with respect to the QSO.
The observations were all taken in seeing-limited mode with $0.125$\arcsec\ pixels,
in  moderate (optical) seeing conditions (1\arcsec--1.5\arcsec), yielding a point spread function (PSF)
of 0.8\arcsec--1.2\arcsec\ in the J-band.

The data reduction is based on pairwise subtraction of the frames 
 as in \citet{ForsterSchreiberN_06a} using the MPE SINFONI pipeline \citep[SPRED,][and references therein]{AbuterR_06a}
and the OH sky line removal scheme of \citet{DaviesR_06a}.
 The   wavelength solution is obtained using the Ar lamp frames, and the absolute wavelength
 is obtained directly from the OH sky emission lines.
Flux calibration is performed using telluric standard stars.

\section{Results}
\label{section:results}

\begin{figure}
%\centering
\includegraphics[width=12cm,angle=-90]{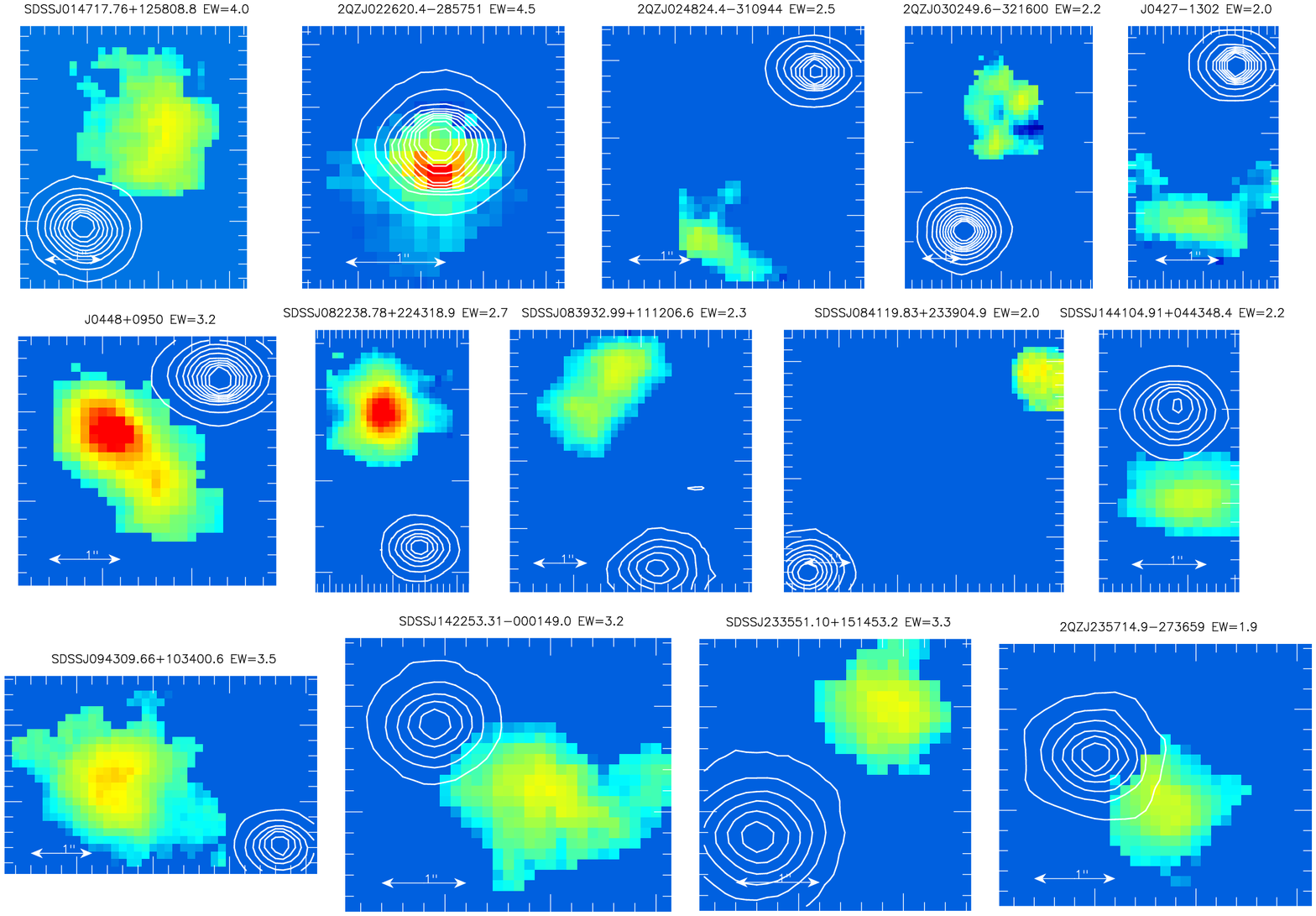}
%\plotone{f1.eps}
\caption{We show the \ndetected\ QSO fields  (sorted by RA.) where the \MgII\ host galaxy was detected via its \Ha\ emission
using the IFU SINFONI. The contours show the QSO continuum (showing the field seeing). The arrow represents 1" ($\sim8$~kpc at $z\simeq1$)
and the impact parameters range from 0.2 to 4". 
One galaxy is seen 0.2" away from the QSO (2QZJ0226-28), well inside the seeing disk (0.8") shown by the QSO continuum contours,
  showing the power of untangling spatially overlapping light when using the full 2-dimensional spectral information from IFU data.
North is up, East is left, color-coding reflects the relative \Ha\ fluxes of the sources. \label{fig:example}}
\end{figure}

From the SINFONI data-cubes, we detect an \Ha\ emitter within $\pm 200$~\kms\ of the predicted
location from the \MgII\ redshift  in \ndetected\ of the \nfield\ fields
down to a flux limit of 4$\times10^{-18}$~\flux\ (1-$\sigma$) (see Table~1).
In other words,  we have a 67\%\ detection rate.
The undetected hosts are either fainter than our flux limit or outside the field of view.
Fig.~\ref{fig:example}  shows the \Ha\ flux maps for all of the  detected host galaxies.
 In all frames, the QSO PSF (typically 0.8-1.2\arcsec\ FWHM) was subtracted from the \Ha\ flux maps and is shown as contours.

As seen from Fig.~\ref{fig:example}, 
the impact parameters range from 0.2\arcsec\ to 4\arcsec\ (1 to 25~kpc) with one at $\sim 6.7$" (J0841+2339). 
The very strong absorber (\EW=4.5\AA) towards the QSO 2QZJ0226-28
 is  detected with an impact parameter of 0.2\arcsec, well within the seeing disk. 
 We can rule out the possibility of this emission line being 
from the QSO. Indeed, the emission line at 1.327$\mu$m  corresponds to   4185\AA\
in the rest-frame of the QSO and therefore does not correspond to any known atomic transition.
Furthermore, the emission line is clearly spatially offset from the QSO centroid.
This shows that it is possible to decouple the intervening galaxy and the QSO
owing to the information contained in the IFU data cube. 
 
It is extremely unlikely that all the \ndetected\ emission lines shown in Fig.~\ref{fig:example}
be something else  in relation to other intervening systems along the line of sight. 
For the strongest emitters (J0448,J0302,J0822,J0943), 
we do also detect [NII] $\lambda$6583 and [SII] $\lambda\lambda$6717,6731,
therefore  lifting any ambiquity.
In the cases where only one line  is detected, we can reject 
 the other possibilities (\OII, \OIII, \Hb)   for the following reasons.
[\OIII] $\lambda$5007 is a doublet easily identifiable at the SINFONI resolution $R\sim2200$.
[\OII] $\lambda$3727 would not be resolved, but that would put the galaxy 
at a redshift higher than the QSO emission redshift.
In most cases, the same argument applies for \Hb. When not,
 we have checked the QSO spectrum for possible corresponding absorbers (\MgII,\FeII)
 and found nothing in the QSO spectrum at the corresponding location.

After more than two decades, only $<100$ absorption-selected galaxies have been identified in emission, and
 mostly at $z\simeq1$ \citep[e.g.][and references therein]{KulkarniV_06a} using broad band imaging \citep[e.g.][]{LeBrunV_97a,ChenH-W_03a}.
Low-redshift ($z<1$) \MgII\ absorbers have been identified more easily  \citep{BergeronJ_91a,SteidelC_94a,LeBrunV_97a}
than low-$z$ and high-$z$ damped \lya\ absorbers (DLAs), which are all also \MgII\ absorbers \citep{RaoS_06a}.    
Our  67\%\   detection rate of the host \Ha\ emission
 is thus in contrast to most previous \Ha\ studies
of absorption-selected galaxies since our sample of strong \MgII\ absorbers ($\EW>2$~\AA) is dominated by DLA `candidates'
meeting the criterion of \citet{RaoS_06a} (see Table~1) and is due to several factors. 
If our careful avoidance of OH sky emission lines plays a significant role,
it is  the SINFONI throughput that enabled us to reach
flux limits unavailable previously from the ground, reaching $1.2\times10^{-17}$~\flux\ or a SFR of $\sim$0.5~\mpy at $z=1$ in less than 1~hr.
For comparison, previous \Ha\ surveys for $z>2$ DLAs  include the ground-based surveys of (e.g.)
 \citet{MannucciF_98a} unveiling several candidates down to $1.4\times10^{-16}$\flux($3\sigma$), 
 \citet{vanderWerf_00a} unveiling several candidates down to $1.7\times10^{-16}$\flux($3\sigma$),
 \citet{BunkerA_99a} showing no detection down to $7$--$16\times10^{-17}$\flux ($3\sigma$),  
 the {\it HST}/NICMOS  survey of the $z=0.656$ DLA (\EW=1.5~\AA) by 
  \citet{BoucheN_01a}  showing no detections down to $3.7\times10^{-17}$\flux ($3\sigma$), 
and the NICMOS   survey of a $z=1.89$ DLA by \citet{KulkarniV_01a} showing no detection down to $1.4\times10^{-17}$\flux (3$\sigma$).

The large number of \Ha\ detections in this survey shows the relatively high star formation rates
of the host galaxy given our shallow exposures. 
The averaged integrated \Ha\ luminosity for our sample is $L_{\Ha}\simeq 6.4\times10^{41}$~erg~s$^{-1}$ (median $4.2\times10^{41}$~erg~s$^{-1}$),
 uncorrected for extinction. This is  about $>4$~times that of the starburst M82 whose raw $L_{\Ha}$ is $1.5\times10^{41}$~erg~s$^{-1}$ \citep{LehnertM_99a}.
If we adopt a statistical correction of $A_V=0.8$mag as derived
      for local star-forming and starburst galaxies of similar
      \Ha\ luminosities \citep{KennicuttR_98a,BuatV_02a},
      we infer an average (median) SFR of $\sim$12~(7)~\mpy\ using the \citet{KennicuttR_98a} flux conversion
 for a Salpeter IMF, and $\sim$7~(4)~\mpy\ using the \citet{ChabrierG_03a} IMF (both from 0.1~\msun\ to 100~\msun). 
  Table~\ref{table:summary} lists the derived properties of our sample.

\begin{figure}
%\centering
%\includegraphics[width=8cm]{fig2.eps}
\plotone{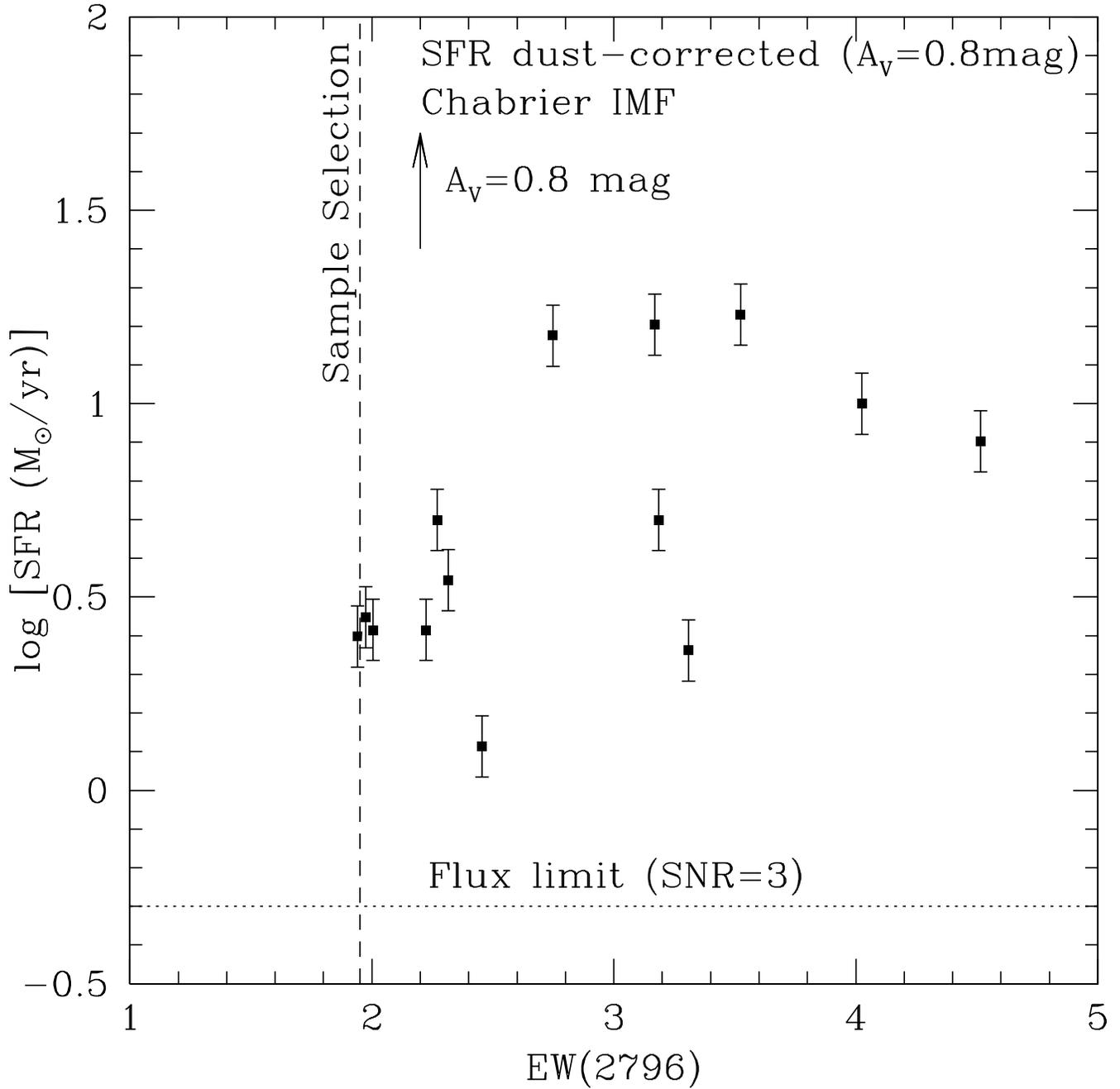}
\caption{The star-formation rate (SFR) from the \Ha\ emission of \MgII\ hosts
as a function of rest-frame equivalent width \EW. SFR and \EW\ may be correlated, indicating
a physical connection between the host starburst and the gas seen in absorption.
The cut at $\EW=2$~\AA\ is due to our sample selection. Our SFR flux limit (SNR=3) is shown as the dotted line.   \label{fig:SFR}}
\end{figure}

Fig.~\ref{fig:SFR} shows the SFR as a function of \EW.
It appears that there is a correlation between the SFR and \EW,
which is significant at more than $2$-$\sigma$: 
the Pearson correlation coefficient is $\rho = 0.59$ with a $P$-value$=0.024\;<0.05$, and a Spearman rank test gives $\rho=0.53$ and $P=0.048$.
 The corrrelation is very consistent with the analysis of \citet{ZibettiS_07a} who found that the hosts of the strongest \MgII\ absorbers are bluer. 
If stronger starbursts produce larger outflows (whose signature is more clouds over a larger $\Delta v$),
such a correlation would naturally arise in the outflow scenario. We will attempt to model this
in more detail  in Bouch\'e et al. (in preparation).

The inferred SFRs are moderate, but whether or not the \Ha\ emitters seen in Fig.~\ref{fig:example} are producing winds depends on the SFR
per unit area, $\Sigma$, not on the integrated SFR. For instance, the starbursts M82 \citep[SFR$\sim 2$ \mpy,][]{LehnertM_99a} and NGC 1569 
\citep[SFR$\sim1$ \mpy,][]{MartinC_02a}   have $\Sigma$   greater than the 0.1~\mpy~kpc$^{-2}$ threshold  
for harboring SN driven outflows \citep{HeckmanT_03a}, and do show strong outflows. Already, the majority of our sample meet this criteria. Higher spatial resolution data
and better size constraints will enable us to test whether strong \MgII\ selected galaxies always meet the outflow threshold
and, in the cases where it does, to probe the outflow physical properties.

As stated in section~1,  \Ha\ ought to be present and detected
in the starburst--outflow scenario. Had we not detected \Ha, this would have ruled out the outflow scenario.
The  high detection fraction thus prevents us from ruling out the outflow scenario.
The relatively high average SFR (compared to M82) is in fact very consistent with the prediction of the
 starburst--outflow scenario, and we view this result as supporting the outflow scenario,
 but other interpretations are possible.
Note that our conclusion {\it only} applies for
 galaxies selected with the strongest \MgII\ absorption equivalent widths ($\EW\,>\,2$~\AA).
 \MgII\ absorption systems with 
$\EW\,<\,2$~\AA\ are much more numerous and may select a more diverse class of galaxies 
\citep[e.g.][]{BergeronJ_91a,SteidelC_94a}.
%The starburst scenario opens the possibility to study M82-analogs at $z\simeq1 $ and up to $z=2.2$
%when one combines the kinematics in absorption from \MgII\ and in emission.

\begin{figure}
%\centering
%\includegraphics[width=8cm]{fig3.eps}
\plotone{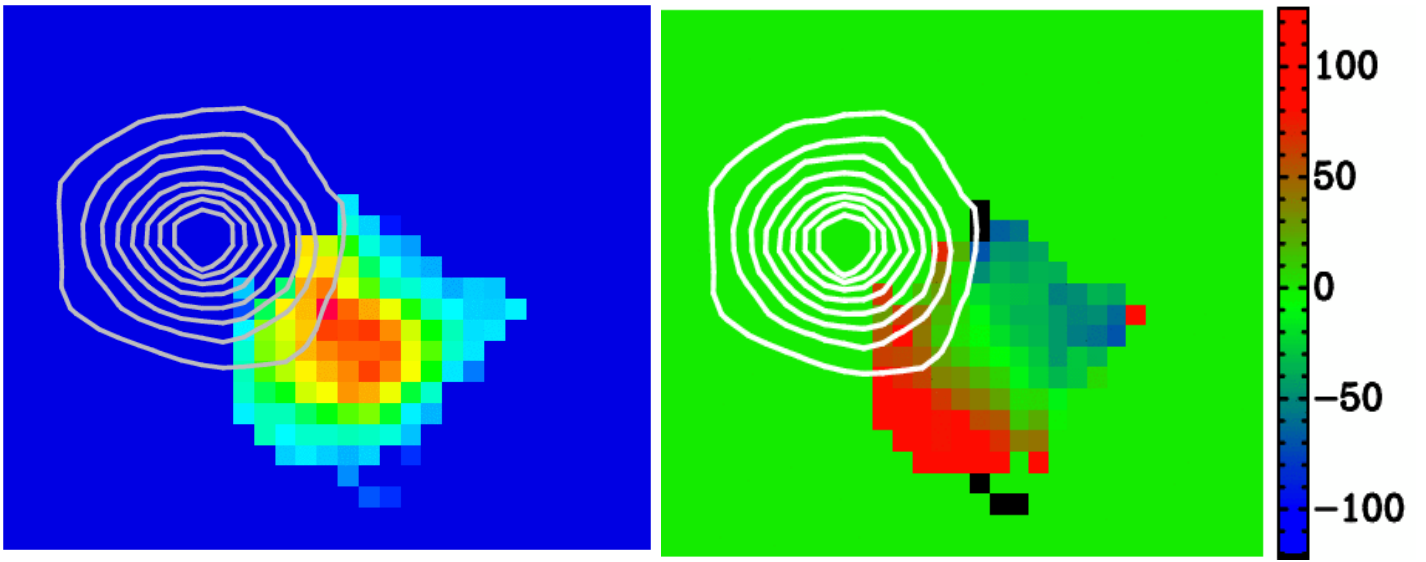}
\caption{Comparison between the \Ha\ emission map (left) and the two dimensional velocity field (right) for one of our \MgII\ host galaxies.
The QSO continuum is shown as grey contours.
The kinematics reveal that the kinematic major axis is perpendicular to the QSO--galaxy vector,
 i.e. the QSO is located along the minor axis, which is further
evidence that the gas traced by \MgII\ is part of the halo gas and not the disk. The maximum rotation velocity is  $V_{\rm max}\sim80$~\kms\
for this galaxy  and is typical for our sample. Using $V_{\rm max}$
as a proxy for the circular velocity, the inferred mean halo mass is $<\log M_h(\msun)>\sim 11.2$ for the sample. \label{fig:mass}}
\end{figure}

Fig.~\ref{fig:mass} shows an example of the two dimensional velocity field for one of the \MgII\ host galaxies.
For the remainder of the sample, when we can extract the velocity field, we find that  the  host galaxy of \MgII\ absorbers with $\EW\;>\;2$~\AA\
reside in halos with masses ranging from $\log M_h(\msun)\sim 9.9$--11.9 with a logarithmic mean
    $<\log M_h(\msun)>\sim 11.2$, once corrected for inclination. This   is 10 times smaller than the Milky Way, and
    in good agreement with the halo masses predicted from the clustering analysis of \citet{BoucheN_04a} and
    \citet{BoucheN_06c}.
   Here we are using the maximum velocity $V_{\rm max}$ as a proxy for the circular velocity.
In N. Bouch\'e et al. (in preparation), we will present the kinematic results in detail.

\acknowledgments

We thank the ESO Paranal staff for their continuous support and the referee for his/her detailed report that led to an improved manuscript. 
 
\bibliographystyle{apj}
%\bibliography{references}

\clearpage

%\begin{figure}
%\centering
%\includegraphics[width=14cm]{figs/mosaic1.Ha-cont.ps}
%\caption{\label{fig:mosaic1}}
%\end{figure}
%\begin{figure}
%\centering
%\includegraphics[width=14cm]{figs/mosaic2.Ha-cont.ps}
%\caption{\label{fig:mosaic2}}
%\end{figure}

\clearpage

\clearpage

\setlength{\tabcolsep}{1pt}

\begin{table}
\begin{tabular}{llllcccccc}
\hline
Sight Line & $z_{\rm qso}$ & $z_{\rm abs}$ & \EW\ & $\log $\NHI\tablenotemark{a} & PSF  &  $b$	&	$\rho$ & $f_{\Ha}$~\tablenotemark{b} & SFR   \\
	   &		&		&(\AA)	& (\cmsq) & (arcsec)& (arcsec)	& (kpc)	& \flux	& \mpy	  \\
\hline
{Sightlines with Detections}\\
\hline
%%YES 14
SDSSJ014717.76+125808.8  &1.5030 &1.03906 &4.025   &1 & 0.9	  &1.9    &  15  & (1.7$\pm0.4$)E-16 & 10    \\ 
2QZJ022620.4-285751 	 &2.171  &1.0208  &4.515   &...& 0.7	  &0.25   &  2   & (1.5$\pm0.3$)E-16 & 8    \\ 
2QZJ024824.4-310944 	 &1.399  &0.7906  &2.455   &...& 0.7	  &3.0    &  24  & (4.7$\pm1.1$)E-17 & 1.3  \\ 
2QZJ030249.6-321600 	 &0.898  &0.8217  &2.27	   &...& 0.9	  &3.5    &  27  & (1.6$\pm0.3$)E-16 & 5.0    \\ 
J042707.3-130253.5	 &2.168  &1.03450 &2.005   &1 & 0.7	  &2.6    &  20  & (4.4$\pm1.0$)E-17 & 2.6   \\
J044821.8+095051.7	 &2.115  &0.83920 &3.169   &1 & 0.8	  &1.8    &  14  & (3.8$\pm0.8$)E-16 & 16   \\ 
SDSSJ082238.78+224318.9  &1.6200 &0.81049 &2.749   &1 & 0.9	  &3.1    &  24  & (4.7$\pm0.9$)E-16 & 15   \\ 
SDSSJ083932.99+111206.6	 &2.696  &0.78740 &2.32    &0 & 1.0	  &3.7    &  29  & (1.2$\pm0.2$)E-16 & 3.5  \\ 
SDSSJ084119.83+233904.9  &1.5080 &0.81193 &1.974   &1 & 0.8	  &6.7    &  54  & (1.0$\pm1.9$)E-16 & 2.8  \\ 
SDSSJ094309.66+103400.6  &1.2390 &0.99646 &3.525   &1 & 0.8	  &3.0    &  24  & (3.3$\pm0.7$)E-16 & 17   \\ 
SDSSJ142253.31-000149.0  &1.0830 &0.90969 &3.185   &1 & 0.8	  &1.5    &  12  & (1.2$\pm0.3$)E-16 & 5.0    \\ 
SDSSJ144104.91+044348.4  &1.1120 &1.03878 &2.22    & 1& 0.7       &1.4    &  11  & (4.5$\pm1.0$)E-17 & 2.6  \\ 
SDSSJ233551.10+151453.2  &0.8920 &0.85568 &3.308   &1 & 1.0	  &2.1    &  17  & (6.5$\pm1.4$)E-17 & 2.3  \\ 
2QZJ235714.9-273659 	 &1.732  &0.8149  &1.940   &...& 0.8	  &1.0    &  8   & (8.6$\pm1.7$)E-17 & 2.5  \\ 
\hline
Sightlines with Non-detections\\
\hline
%%%%NO 6
2QZJ014729.4-272915 	 & 1.697  & 0.811  & 2.17   & ...    & 0.7  &  NA & NA  & $<1.2$E-17 &  $<0.6$  \\
SDSSJ100715.52+004258.4	 & 1.681  & 1.0373 & 2.98   & 21.15  & 1.2  &  NA & NA  & $<1.2$E-17 &  $<0.6$ \\
SDSSJ110744.61+095527.0  &1.222   &0.80229 & 3.87   & 1      & 1.2  &  NA & NA  & $<1.2$E-17 &  $<0.6$ \\
SDSSJ142650.9+005150.5	 &1.333   &0.8424  & 2.61   & 19.65  & 0.7  &  NA & NA  & $<1.2$E-17 &  $<0.6$ \\
J215145.8+213013.5 	 & 1.538  & 1.0023 & 2.46   & 19.30  & 1.2  &  NA & NA  & $<1.2$E-17 &  $<0.6$  \\
SDSSJ211100.20-005218.3  &  1.6860& 1.02094& 3.608  & 1      & 1.1  &  NA & NA  & $<1.2$E-17 &  $<0.6$  \\		
2QZJ232330.4-292123      & 1.547  & 0.811  & 2.0430 & ...    & 1.0  &  NA & NA  & $<1.2$E-17 &  $<0.6$  \\
\hline 
\tablenotetext{a}{a `1' indicates that the system meets the \citet{RaoS_06a} criteria for being a DLA.}
\tablenotetext{b}{the 3-$\sigma$ upper limits for the non-detections are computed for an unresolved source spread over $\sim200$ pixels, spread over 32 spatial pixels and
spectral FWHM=6~pix=9\AA\ (i.e. $R_{1/2}=0.4$"$\simeq3$~kpc, and FWHM=250~\kms). The same dust-correction was applied to the SFR limits.}

\end{tabular}						        
\caption{Sample Properties. \label{table:summary}}
\end{table}

\end{document}